\documentclass[epj]{svjour}
\usepackage{graphics}
\begin{document}
\title{Evidence of stochastic resonance \textbf{in the mating behavior of \emph{Nezara viridula}} (L.)}
\author{S. Spezia\inst{1,}\thanks{\emph{e-mail:} stefano.spezia@gmail.com},
L. Curcio\inst{1}, A. Fiasconaro\inst{1,2}, N. Pizzolato\inst{1}, D. Valenti\inst{1},
B. Spagnolo\inst{1}, P. Lo Bue\inst{3}, E. Peri\inst{3}, S. Colazza\inst{3}
%
}                     
%
%
\institute{Dipartimento di Fisica e Tecnologie Relative, Group of
Interdisciplinary Physics \thanks{\emph{URL:}
http://gip.dft.unipa.it}, Universit\`a di Palermo, Viale delle
Scienze pad.~18, I-90128 Palermo, Italy \and Mark Kac Complex
Systems Research Center and Marian~Smoluchowski Institute of
Physics, Jagellonian University, ul. Reymonta 4, 30-059 Krak\'ow,
Poland \and Dipartimento di Scienze Entomologiche, Fitopatologiche,
Microbiologiche agrarie e Zootecniche, Universit\`a di Palermo,
Viale delle Scienze pad.~5, I-90128 Palermo, Italy}
\date{Received: date / Revised version: date}
%
\abstract{We investigate the role of the noise in the mating
behavior between individuals of \emph{Nezara viridula} (L.), by
analyzing the temporal and spectral features of the non-pulsed type
female calling song emitted by single individuals. We have measured
the threshold level for the signal detection, by performing
experiments with the calling signal at different intensities and
analyzing the insect response by directionality tests performed on a
group of male individuals. By using a sub-threshold signal and an
acoustic Gaussian noise source, we have investigated the insect
response for different levels of noise, finding behavioral
activation for suitable noise intensities. In particular, the
percentage of insects which react to the sub-threshold signal, shows
a non-monotonic behavior, characterized by the presence of a
maximum, for increasing levels of the noise intensity. This
constructive interplay between external noise and calling signal is
the signature of the non-dynamical stochastic resonance phenomenon.
Finally, we describe the behavioral activation statistics by a soft
threshold model which shows stochastic resonance. We find that the
maximum of the ensemble average of the input-output
cross-correlation occurs at a value of the noise intensity very
close to that for which the behavioral response has a maximum.
\PACS{
      {87.18.Tt}{Noise in biological systems} \and
      {87.50.yg}{Biophysical mechanisms of interaction} \and
      {05.40.-a}{Stochastic processes}
} 
} 
\authorrunning{\texttt{S. Spezia et al}}
\titlerunning{\texttt{Stochastic resonance in \emph{Nezara viridula} mating behavior}}

\maketitle
\section{Introduction}
\label{sec:1} In the last twenty years several experimental and
theoretical investigations have been carried out on noise-induced
effects in neuronal dynamics, in excitable systems and in threshold
physical
systems~\cite{Bra94,Mos94,Gin95,Pei95,Pik97,Noz98,Lon98,Sto00,Wie94,Gam95,Wan00,Lin04}.
In particular, resonant activation, stochastic resonance and noise
enhanced stability phenomena in neuronal activation have been
recently discussed~\cite{Lin04,Pol05,Dua08}.

The functionality of a complex biological system depends on the
correct exchange of information between the component parts. In
natural systems the environmental noise always affects the signal
that carries the information. Usually high levels of noise make
difficult to reveal signals, so that in everyday life the noise is
generally considered harmful in detecting and transferring
information. On the other hand, nature consists of open systems
characterized by interactions which are (i) inherently non-linear
and (ii) noisy, due to the influence of the
environment~\cite{Spa04}. Under specific conditions, the noise can
constructively interacts with the system, so that effects induced by
the noise, such as stochastic resonance (SR), can improve the
conditions for signal detection.

Stochastic resonance (SR), initially observed in the temperature
cycles of the Earth \cite{Benzi}, is a counterintuitive phenomenon,
whereby the addition of noise to a weak periodic signal causes it to
become detectable or enhances the amount of transmitted information
through the
system~\cite{Mos94,Gin95,Pei95,Noz98,Lon98,Sto00,Wie94,Gam95,Wan00,%
Lin04,Vil98,Dou93,Rus99,Fre02,Gre00,Lon91,Bul91,Nei02,Bah02,Gam98,Man94,Gai97}.
When SR occurs, the response of the system undergoes resonance-like
behavior as a function of the noise level. In spite of the fact that
initially this phenomenon was restricted to bistable systems, it is
well known that SR appears in monostable, excitable, and
non-dynamical systems.

Non-dynamical stochastic resonance refers to a situation where the
mere addition of noise can improve the system sensitivity to
discriminate weak information-carrying
signals~\cite{Mos94,Gin95,Vil98}. The age of SR in biology started
in the early 1990s with benchmark publications wherein SR was
revealed in sensory neurons affected by external noise. In
particular, Moss and collaborators set up an experiment to study the
neural response of mechanoreceptor cells of crayfish~\cite{Dou93},
and the enhancement of electrosensory information in paddlefish for
prey capture~\cite{Rus99,Fre02,Gre00}. Such sensory neurons are
ideally suited to exhibit SR as they are intrinsically noisy and
operate as threshold systems~\cite{Lon91,Bul91,Nei02,Bah02}.

In this paper we report on experiments conducted on the response of
\emph{Nezara viridula} (L.) (Heteroptera Pentatomidae) individuals
to sub-threshold signals. Specifically we investigate the role
played by the noise in the communication between individuals of
opposite sex of \emph{N. viridula} through the recognition of
mechanical vibrations transmitted in the
substrate~\cite{Cok99,Cok03,Cok07}. \emph{Nezara viridula}, the
southern green stink bug, is a cosmopolitan insect, occurring
throughout tropical and subtropical regions of Europe, Asia, Africa
and America. This species is highly poly-phagous and it's one of the
most important pentatomid insect pests in the world
\cite{Tod89,Pan00}. \emph{Nezara viridula} has up to five
generations per year~\cite{Bor87,Kir64,Fuc03}.

The mating behavior of \emph{N. viridula} can be divided into
long-range attraction and short-range courtship. The first one
includes those components of the behavior that lead to the arrival
of females in the vicinity of males. The long range attraction
mediated by male attractant pheromone enables both sexes to reach
the same plant. Short-range courtship includes those components that
coordinate the interaction of both sexes once they are in strong
proximity. In this last condition the acoustic stimuli (improperly
called songs) have an important role in the sexual communication
between male and female individuals~\cite{Cok99}.

The sound generating organ is the tymbal organ across the back,
present in adult bugs~\cite{Cok03}. These animals produce vibrations
at the frequency of about $100$ Hz. These vibrations are transmitted
through the legs into the plant stem and detected by vibro-receptors
sited in the legs of the receiving bug~\cite{Bag08}. Because of its
essential role during the mating behavior, the reception of these
signals has been studied in \emph{N. viridula} populations from
Slovenia, Florida, Japan and Australia~\cite{Cok00}.

Due to the importance of acoustic communication in mating behavior,
a possible control of \emph{N. viridula} populations can be achieved
by devising traps that work emitting vibratory signals. In this
context, because of the strong interaction between species and
environment, the role played by external noise in acoustic
communication becomes relevant.

In section~\ref{sec:2} we describe the experimental equipment and
methods used in measurements of the mechanical vibrations emitted by
\emph{N. viridula} individuals. In order to find a threshold level
for the insect behavior we have performed directionality tests on
\emph{N. viridula} male bugs by using different amplitudes of
calling signal. In the presence of a sub-threshold signal mixed with
an external noise, we have investigated the insect response as a
function of different noise intensities. The results, reported in
section~\ref{sec:3}, suggest the presence of noise-induced neuronal
activation for a sub-threshold signal. Therefore, in section
~\ref{sec:4} we discuss the experimental results within the
framework of the soft threshold model which shows stochastic
resonance (TSR) phenomenon. In the final section we draw our
conclusions.

\section{Materials and methods}

\label{sec:2} All experiments have been performed by using \emph{N.
viridula} collected in fields around Palermo, and reared in
laboratory condition~\cite{Col04}. Adult males have been tested at
least ten days after the final moult to ensure their sexual maturity
and a period of isolation of three days from the other
sex~\cite{Cok07,Cok00}.

The vibratory signal of our interest is the sexual calling song of
female sex. This signal has been recorded by the membrane of a conic
low-middle frequency loudspeaker (MONACOR SPH 165 C CARBON with a
diameter of $16.5 \thinspace cm$) and stored in a computer for sound
analysis. For the acquisition, processing and analysis of the sounds
a commercial software has been used. The speaker has been used as
"inverse" microphone, namely an acoustic-electric transducer: the
sounds have been recorded from the non-resonant membrane of a
speaker, carefully chosen to get a good frequency response starting
from $20 \thinspace Hz$. The sound acquisitions have been made
inside an anechoic chamber (sound insulated) at $22-26^\circ$C,
$70-80\%$ of relative moisture and in presence of artificial light.
The choice of this recording set-up has been decided after a
comparative analysis with a recording system based on the use of a
stethoscope. In particular, the speaker membrane shows greater
sensitivity at medium-low frequencies, which are crucial in our
experiment.

The sound has been sampled from the analogic signal source ($44100$
samples per second at $16$-bit) and then it has been filtered by a
$18^{th}$ order Tchebychev filter (type I) with band-pass from $60$
to $400$ Hz. This filtering has been done to cut: (i) the low
frequencies due to the electric network ($50 Hz$) and the conic
loudspeaker, and (ii) the high frequencies due to the electronic
apparatus. Spectral and temporal properties of the measured
non-pulsed female calling songs (NPFCS) have been compared with
those of North America, observing that the sounds produced by adults
of \emph{N. viridula} collected in Sicily have similar spectrum of
that produced by \emph{N. viridula} adults collected in USA with a
slightly different frequency range \cite{Cok07,Cok00,Cok05}.

\begin{figure*}
\begin{center}
\resizebox{2.00\columnwidth}{!}{%
\includegraphics{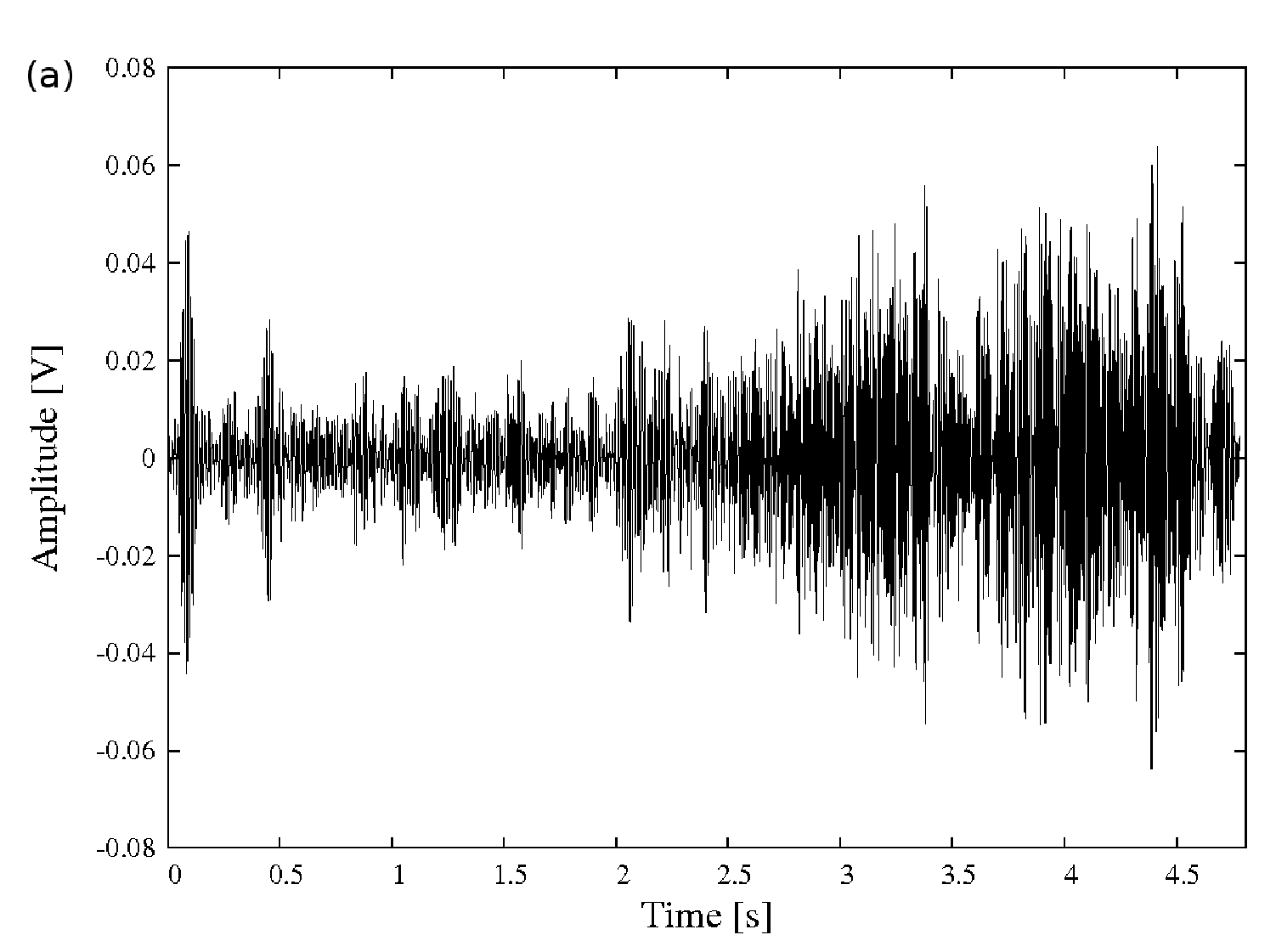}
\includegraphics{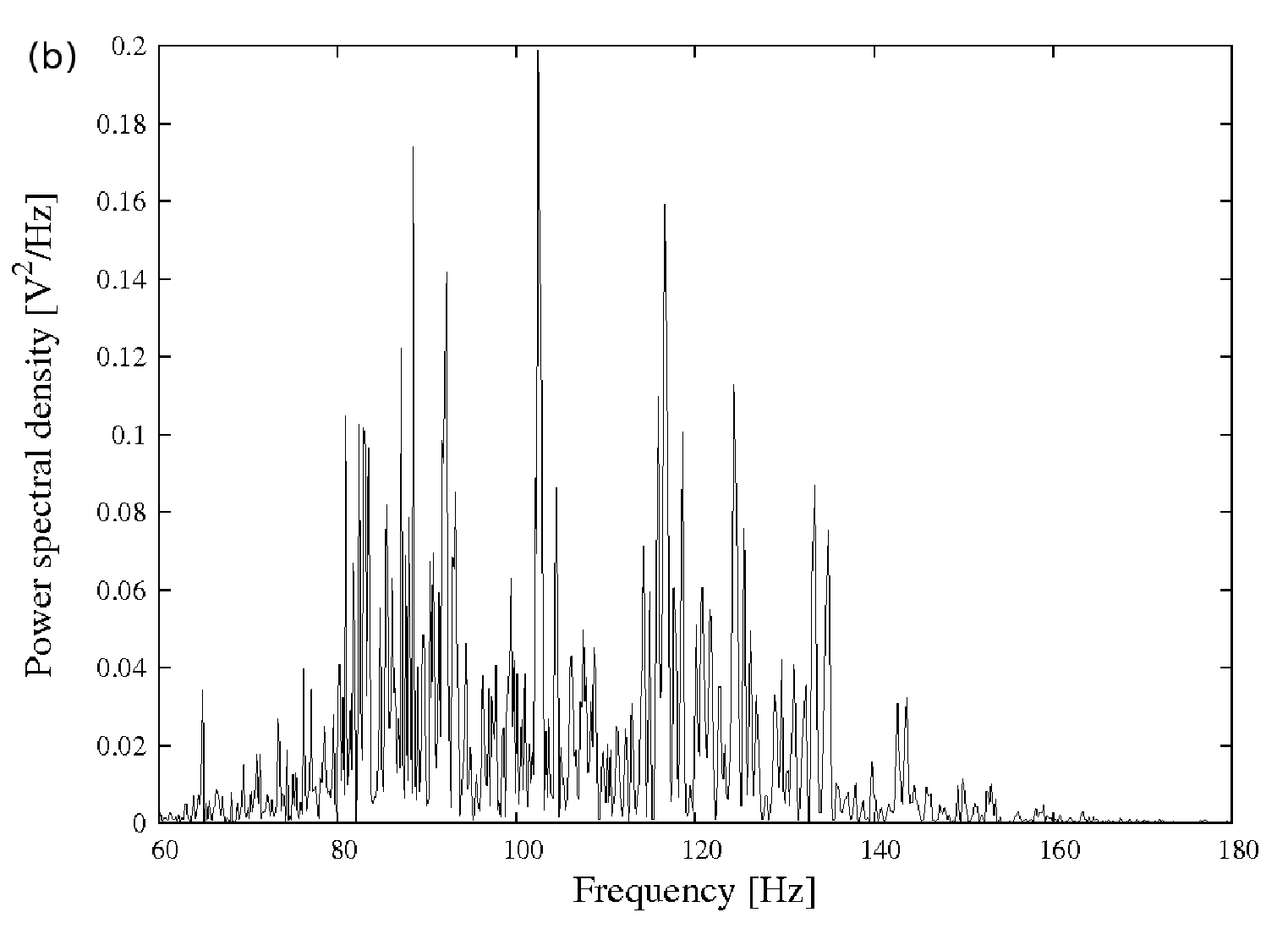}
}
\resizebox{2.00\columnwidth}{!}{%
\includegraphics{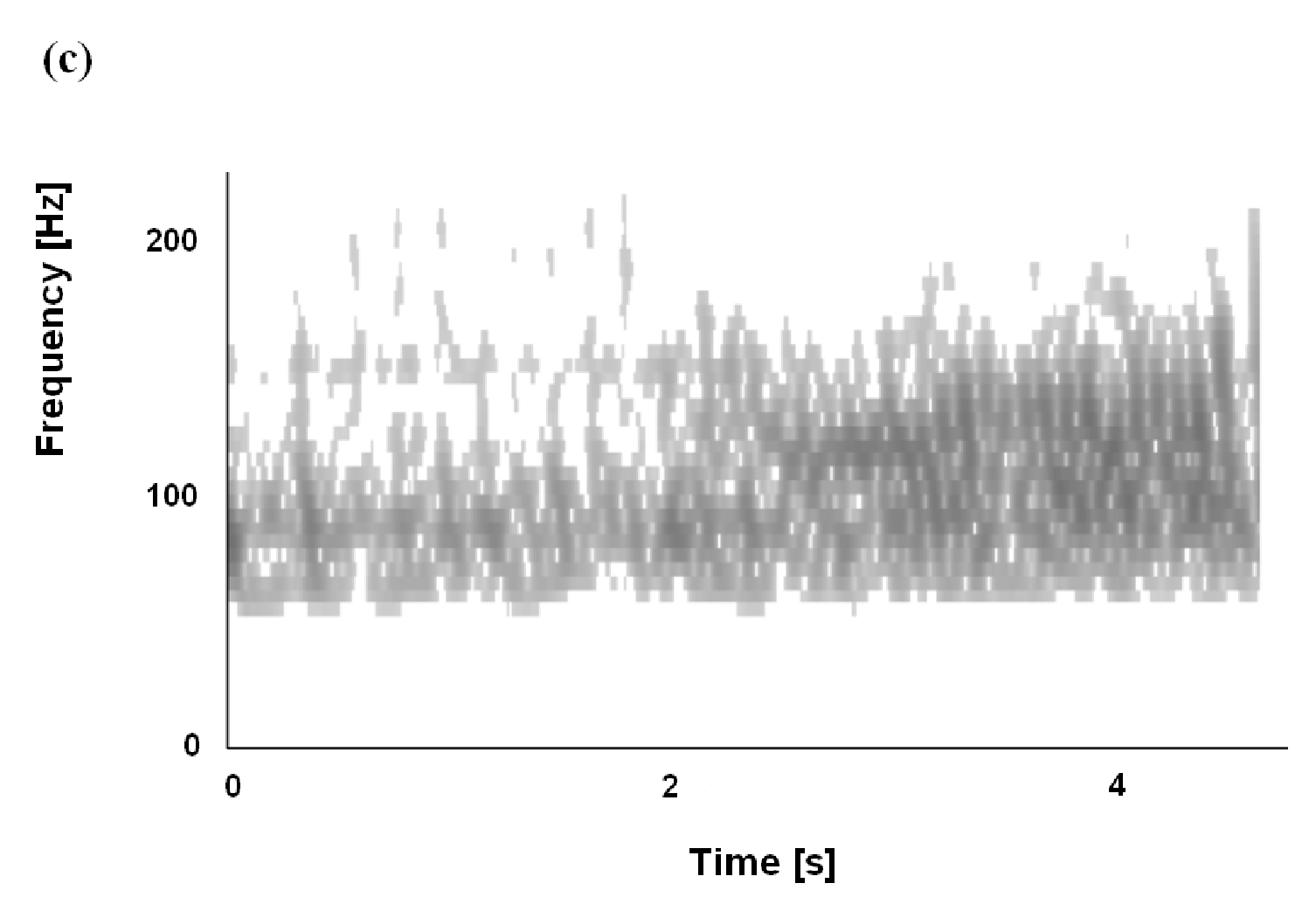}
\includegraphics{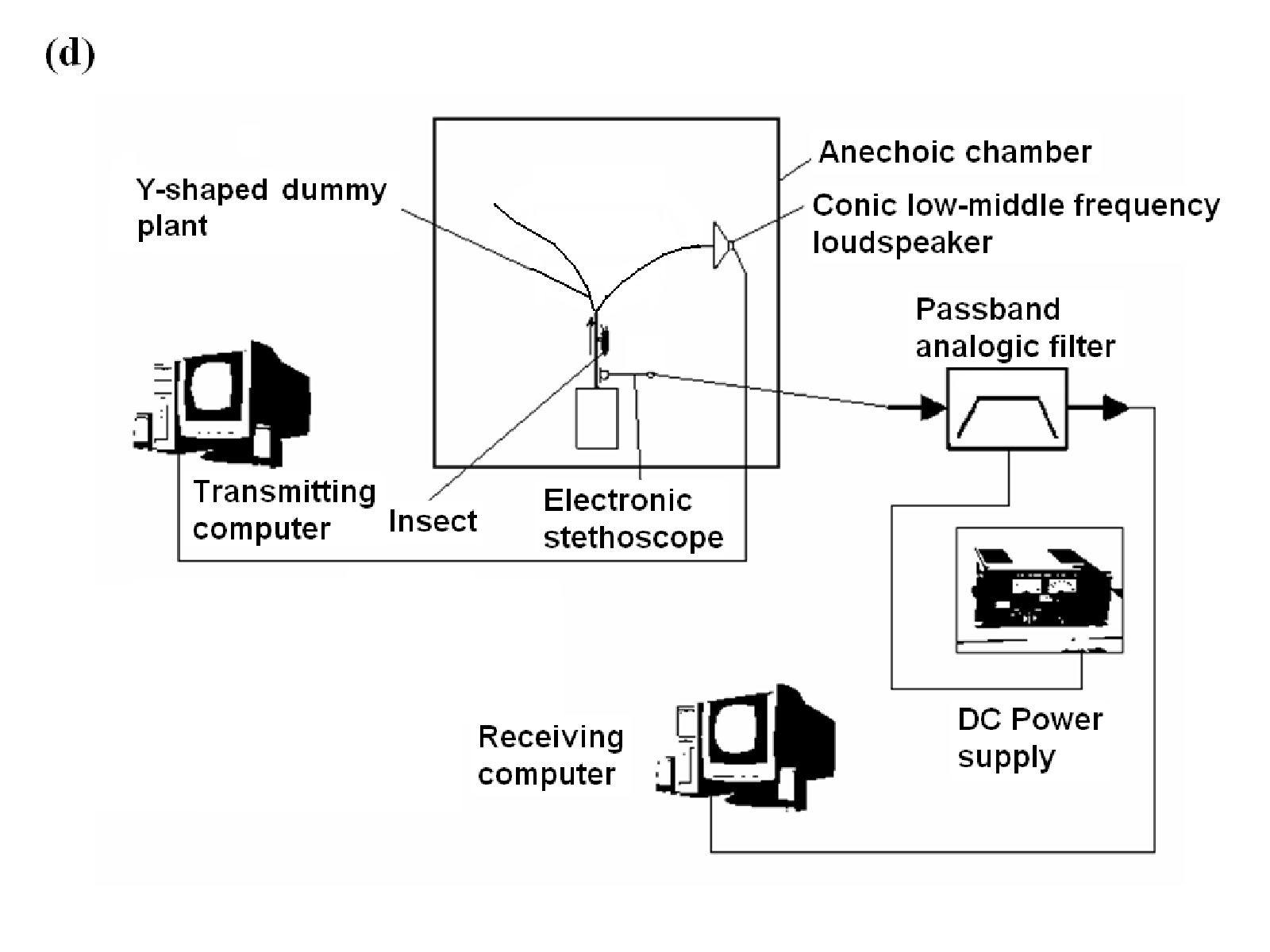}
}
\caption{(a) Oscillogram (b) Power spectrum (c) Sonagram of the non
pulsed type of \emph{Nezara viridula} female calling song; (d) The
block diagram of the experimental setup.}
\label{fig:1}       
\end{center}
\end{figure*}

In Fig.~\ref{fig:1}(a), the oscillogram of NPFCS is shown. The
signal is characterized by a short pre-pulse followed by a longer
one, according to previous experimental findings~\cite{Cok00}. In
Fig.~\ref{fig:1}(b), the power spectrum density (PSD) of NPFCS is
presented. The time length of the NPFCS is $T_s = 4.78$ s. In this
spectrum the dominant frequencies range from $70$ to $170$ Hz and
the subdominant peaks do not exceed $400$ Hz. The maximum peak
occurs at $102.5$ Hz. In Fig.~\ref{fig:1}(c) we report the relative
sonagram, achieved by the Short Time Fourier Transform (STFT)
method. The STFT maps a signal providing information both about
frequencies and occurrence times. It shows that during the first two
seconds (short pre-pulse), the dominant frequency interval is
narrower than the range observed in the successive time window. In
particular in the first time interval the highest frequency doesn't
cross $130$ Hz, whereas in the final one it reaches almost $170$ Hz.

Afterwards, in view of investigating the role of the noise in the
\emph{N. viridula} mating behavior, we have designed an experimental
setup to analyze the ability of male individuals to locate the
source of vibratory signals generated by female insects. In
particular, in order to perform directionality tests, we have
prepared an Y-shaped dummy plant inside an anechoic chamber. This
Y-shaped substrate is constituted by a wood vertical stem, 10 cm
long, and 0.8 - 0.9 cm thick at the top of which there are two wood
branches 25 cm long, and 0.4 cm thick, as shown in
Fig.~\ref{fig:1}(d). The angle between two branches is
$30^\circ-50^\circ$.

The experiment consists of sending a signal on one branch of the
Y-shaped dummy plant and observing the behavior of single male
individuals initially placed at the center of the vertical
stem~\cite{Cok99}. In Fig.~\ref{fig:1}(d) the block diagram of our
experimental set-up is shown. We have sent the vibrational signal at
the right apex of Y-shaped plant. The two lateral branches are not
in contact with the vertical stem. The distance between each branch
and the vertical stem is 0.3 cm. At this stage of the experiment the
cone has been used as electro-acoustic transducer. A response to our
test is achieved when the bug, before choosing one direction in the
Y-shaped structure, has touched the lower end of both branches.
According to this procedure, we have performed more trials for each
fixed intensity (see section 3 for details), counting and recording
the left and right choices of the insects. We have observed the
behavior of the insects in the cross point for different intensities
of the mechanical stimulus. The obtained statistical data on the
directionality choices have been used to determine the intensity
threshold value at which the bugs start to "hear" the calling song.

\section{Experimental results}

\label{sec:3} 
\begin{figure*}[hbt]
\begin{center}
\resizebox{2.10\columnwidth}{!}{%
\includegraphics{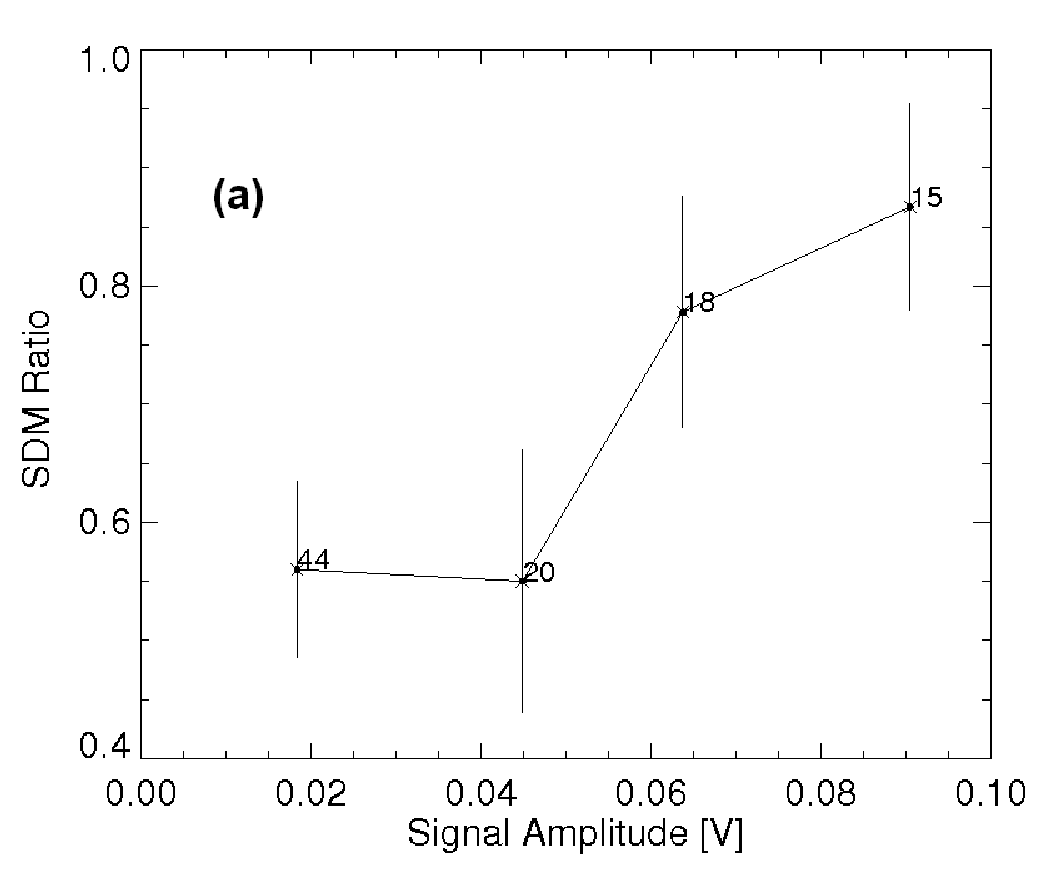}
\includegraphics{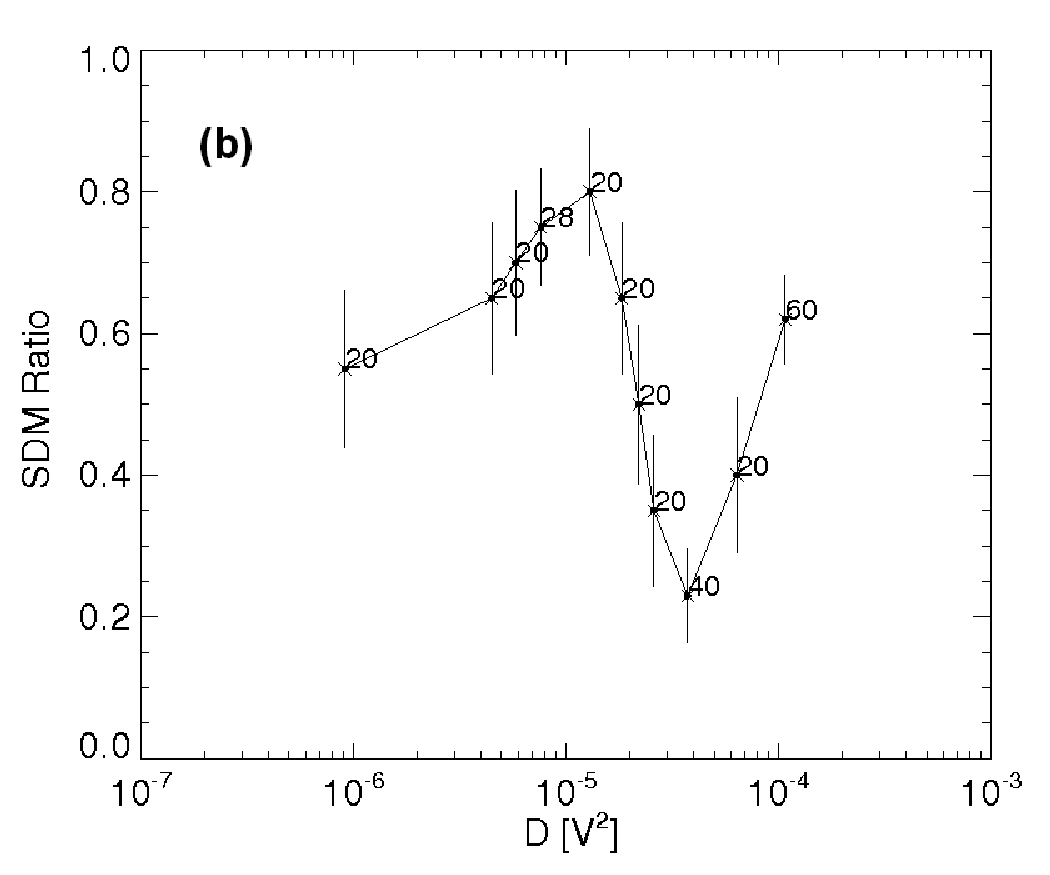}
}
\caption{Plots of the \emph{Source-Direction Movement (SDM) Ratio}
as a function of: (a) the female calling song power (purely
deterministic signal); (b) the noise intensity \emph{D}. In each
experimental value is reported the error bar and beside the
corresponding number of the performed trials.}
\label{fig:2}       
\end{center}
\end{figure*}

The presence of an "oriented" behavior, namely that
the insects tend to choose the branch with the signal source, is
revealed by performing directionality tests on a group of male
individuals. When we observe a percentage of insects higher than
$65\%$ going towards the acoustic source, \emph{Source-Direction
Movement} (SDM), we consider that the signal has been revealed by
the insects. In Fig.~\ref{fig:2}(a) we plot the relative frequency
of SDMs, that is the number of SDMs divided by the total trials, at
different signal intensities. The exact number of trials, performed
for each intensity, is reported beside the corresponding point in
the graph. For every trial actually we used one insect at a time.
Specifically for the experiments related to Fig. 2(a) we used $97$
individuals, of which $63$ have moved towards the sound source. For
the experiments related to Fig. 2(b) we used $288$ animals, of which
$159$ have shown the SR phenomenon. For small values (lower than
$0.044945$ V) of the signal power, a percentage approximately
corresponding to the $50\%$ of the insects chose one direction, the
remaining $50\%$ the other one.
%
Conversely, for values greater than 0.063786 V, the insects show a
preferential behaviour, choosing the direction from which the signal
originates in the $80\%$ of the trials or more. Because of this, we
have chosen the value $0.045$ V of the signal amplitude as the
\emph{threshold level} for the signal detection.

Then, by using a sub-threshold signal plus a Gaussian "white" noise
we have investigated the response of the test insect for different
levels of noise intensity $D$. The "white" noise that we used in our
experiments, to simulate the environmental noise, is generated
numerically by software Matlab. This noise signal has a flat
spectrum up to $22050$ Hz, with a correlation time of about $45$
$\mu$s. However our cone reproduces the sound up to $20$ kHz. This
means that our cut-off frequency is $20$ kHz, which is more than two
order of magnitude greater than the highest frequency emitted by
adults of \emph{N. viridula}. In other words the insect "feels"
white noise. In Fig.~\ref{fig:2}b we report the percentage of SDMs
as a function of $D$, finding the optimal noise intensity that
maximizes the recognition between individuals of opposite sex. The
graph shows a maximum for $D\approx 1.30 \cdot 10^{-5}$ $V^{2}$. For
values of $D>2.50 \cdot 10^{-5}$ $V^{2}$ the percentage of
individuals going towards the acoustic source decreases below $50\%$
reaching $20\%$ for $D\approx 3.75 \cdot 10^{-5}$ $V^{2}$. The other
values of the SDM ratio close to $50\%$, indicate that the
individuals of \emph{N. viridula} have chosen randomly the direction
of their motion, that is no oriented behaviour occurs. The
non-monotonic behaviour of SDM, with a maximum at $D\approx 1.30
\cdot 10^{-5}$ $V^{2}$, indicates that in the presence of a
sub-threshold deterministic signal, the environmental noise can play
a constructive role, amplifying the weak input signal and
contributing to improve the communication among individuals of
\emph{N. viridula}. The occurrence of a minimum in the SDM behaviour
at $D\approx 3.75 \cdot 10^{-5}$ $V^{2}$, will be subject of further
investigations. One possible conjectural explanation is as follows:
when the noise intensity is so great that the signal received from
the vibro-receptors is significantly modified, the male insects are
unable to recognize the female calling song, and they could exchange
it for the song of some rivals.

\indent A further increase of the noise intensity causes the
spectrum of the received signal to become indistinguishable from a
pure environmental noise and therefore the insect is unable to
recognize any signal of \emph{N. viridula} individuals. This implies
that no significative response is observed in terms of percentage of
source-direction movements (SDMs $\sim 50\%$). The non-monotonic
behaviour of SDM, as a function of the noise intensity (see
Fig.~\ref{fig:2}(b), can be considered the signature of the
threshold stochastic resonance.

\section{Threshold Stochastic Resonance}
\label{sec:4} The presence of a maximum in the behaviour of SDM
percentage \emph{vs} $D$ in Fig.~\ref{fig:2}(b) can be explained by
the phenomenon of the soft threshold stochastic resonance. Soft
thresholds are ubiquitous in living systems, in particular in
mechanisms of neurons and neural network such as sensory systems. In
fact, biologic systems, under ordinary conditions, usually don't
exhibit Heaviside-type threshold ("hard" threshold) function as
information transmission function, but rather respond to weak
signals gradually~\cite{Gree04}.

\begin{figure}[h]
\begin{center}
\resizebox{1.00\columnwidth}{!}{%
\includegraphics{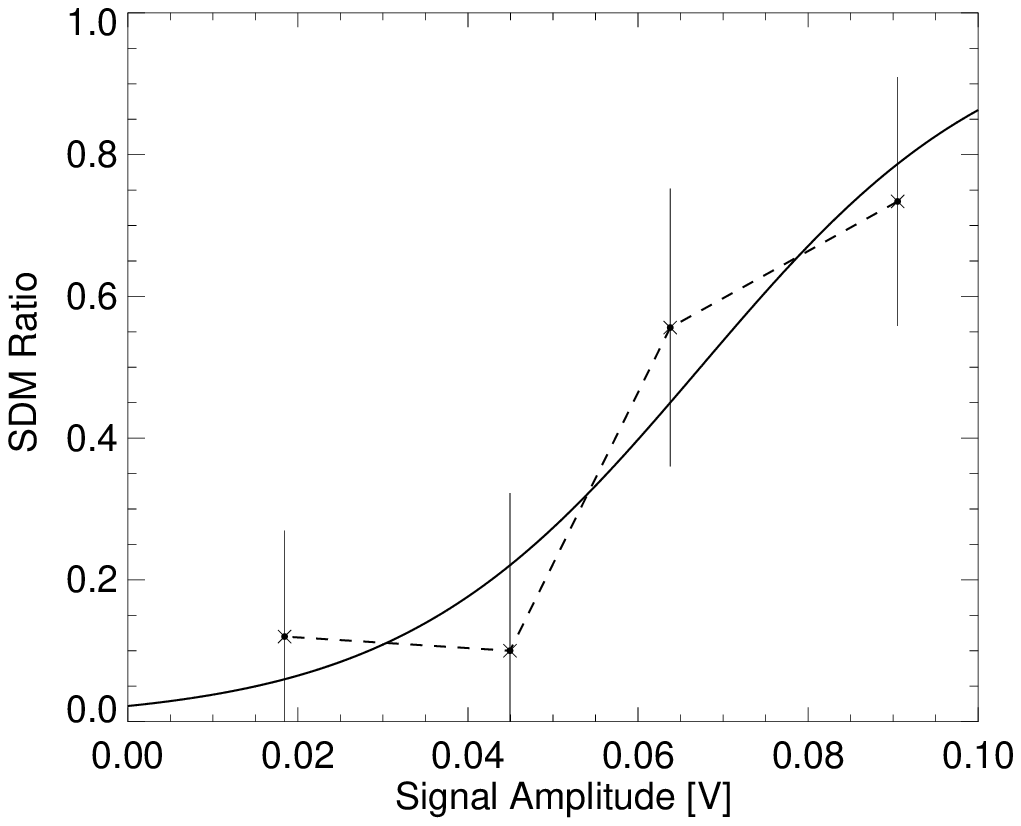}
} \caption{Plot of the logistic function $f(\vartheta)$ versus the
signal amplitude, obtained by a best fitting of the rescaled
experimental data of Fig.~\ref{fig:2}(a).} \label{fig:3}
\end{center}
\end{figure}

The experimental results of Fig.~\ref{fig:2}(a) clearly show that
the individuals of \emph{N. viridula} exhibit "soft" threshold,
which can be represented by a logistic or sigmoid function. In
Fig.~\ref{fig:3} we compare this sigmoid curve with the experimental
data of Fig.~\ref{fig:2}(a), rescaled in such a way that the value
of 50\% of SDM corresponds to zero level. In fact this level
indicate that insects choose randomly the direction of their motion,
without any oriented behaviour. We rescale the experimental values
of the \emph{Source-Direction Movement Ratio} as:
$SDMratio_{rescaled}=2*(SDM ratio(\vartheta)-0.5)$, where
$\vartheta$ is the amplitude of the stimulus, so that $50\%$ of SDM
ratio of Fig.~\ref{fig:2}a corresponds to zero value of the function
\emph{f} in Fig.~\ref{fig:2} and $100\%$ the value $1$ for
$f(\vartheta)$. In human and animal psychophysics, the relationship
between an organism's sensitivity and a sensory stimulus is called
"psychometric function"~\cite{Gree04}. As best fitting of the
experimental data we used as psychometric function the logistic
function

\begin{equation}
f(\vartheta)=\frac{1}{1+a \cdot \exp(-(\vartheta-c)/b)},
\end{equation}
where $a = 2.00464$, $b = 0.0177468$, and $c = 0.055$.

\begin{figure}[h]
\begin{center}
\resizebox{1.00\columnwidth}{!}{%
\includegraphics{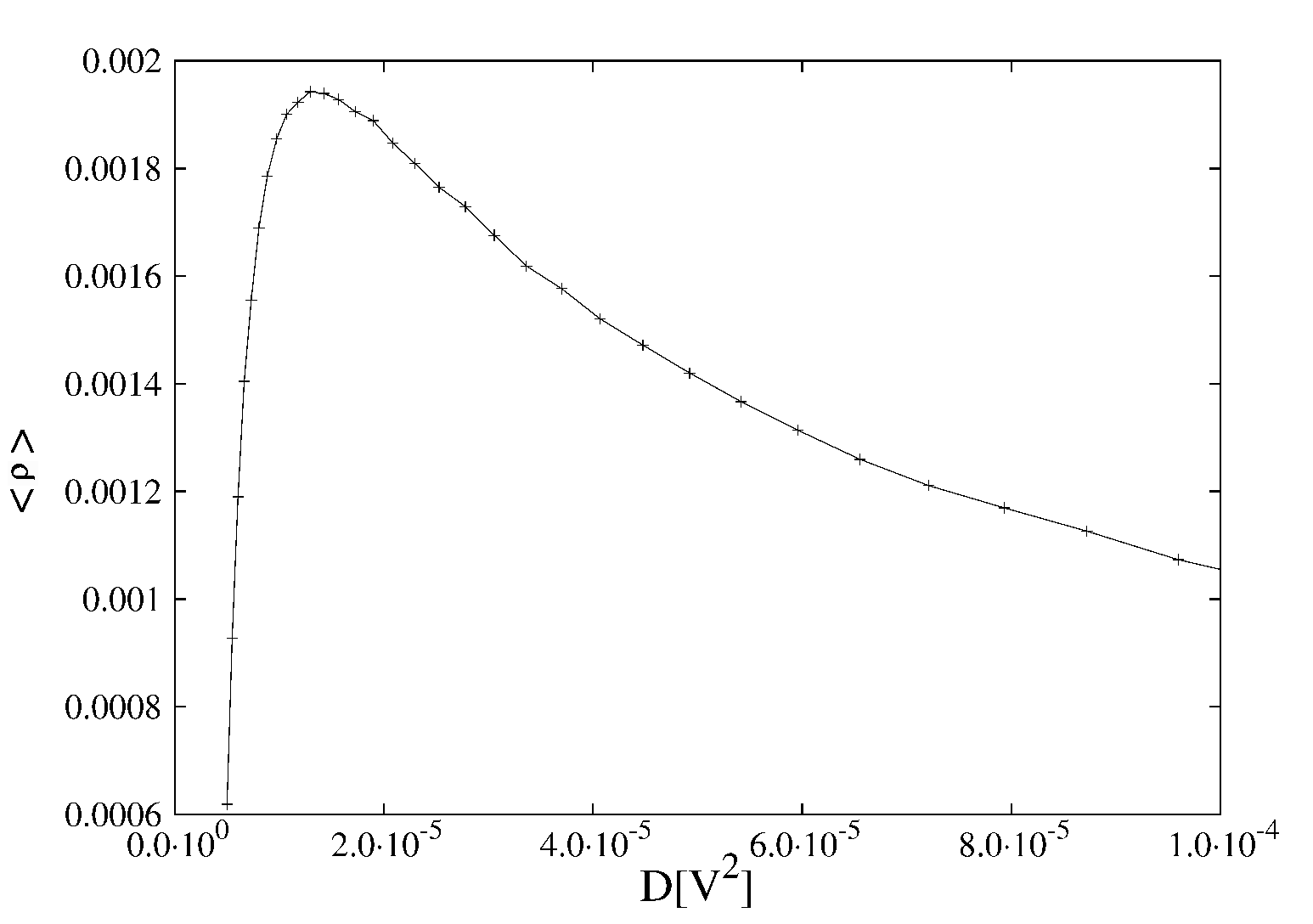}
} \caption{Ensemble average of the input-output cross-correlation
$<\rho>$ versus the noise intensity \emph{D}. The maximum of the
curve is at $D \approx 1.29 \cdot 10^{-5}$ $V^{2}$.} \label{fig:4}
\end{center}
\end{figure}

We have simulated our biological system by a weak signal of maximum
amplitude $0.045$ $V$, obtained by the recorded female calling song
as the input signal, and by using the following model

\begin{eqnarray}
y(t)=\left\{
\begin{array}
[c]{ll}%
s(t) \cdot f(s(t)), &  \qquad s(t)\geq 0\\
0, & \qquad s(t)<0\\
\end{array}
\right .
\label{soft}
\end{eqnarray}
with $s(t) = x(t)+ n(t)$. Here $x(t)$ is the weak input signal,
$n(t)$ is the input noise signal and $y(t)$ is the output signal.
Because of the aperiodicity and the broad spectrum of the input
signal, the signal to noise ratio is not an adequate measure of
stochastic resonance phenomenon. We used a cross-correlation measure
as introduced in Refs.~\cite{Coll95,Dutta}

\begin{equation}
<\rho> = \left < \frac{\overline{x(t) \cdot
y(t)}}{\sqrt{\overline{(x(t)-\overline{x})^{2}} \cdot
\overline{(y(t)-\overline{y})^{2}}}} \right >,
\end{equation}
where the overbar denotes an average over time $T_s$ ($T_s = 4.78$
s.), and the brackets mean ensemble average. The ensemble average of
the input-output cross-correlation coefficient $<\rho>$ as a
function of the noise intensity is reported in Fig.~\ref{fig:4}. For
each value of the noise intensity we have performed $100$ numerical
realizations. The ensemble averaged cross-correlation coefficient
takes its maximum at $D \approx 1.29 \cdot 10^{-5}$ $V^{2}$, which
is very close to the value of the noise intensity that maximizes the
SDM ratio (see Fig.~\ref{fig:2}(b)).

Concerning the role of the internal noise, as reported in the paper
by Gailey et al.~\cite{Gai97}, the peculiar nonmonotonic behaviour
reported in our Fig.~\ref{fig:2}(b) suggests the possibility of the
presence of a second maximum in the curve, for noise intensity
values greater than $10^{-4}$. The presence of this second maximum
indicates that the first one could be ascribed to the internal noise
always present in biological systems. We will investigate in more
detail the behaviour of the SDM Ratio as a function of the external
noise intensity in a forthcoming paper.

The results obtained from our model suggest that in the biological
system analyzed, the stochastic resonance plays a key role, since it
permits to extract information from a weak deterministic signal,
thanks to the constructive action of the environmental noise. In
other words there is a suitable noise intensity which maximizes the
behavioural response of the green bugs and this effect can be
described by a soft threshold model which shows stochastic
resonance.

\section{Conclusions}
\label{sec:6} In this work we have investigated the role of the
noise in the vibrational communications occurring during the mating
of \emph{N. viridula}. In our experiments, by analyzing the response
of the insects to a deterministic signal (calling song), we have
determined the threshold for the neural activation in insect
individuals. By using a sub-threshold signal we have analyzed the
insect response in the presence of an external noise source. We have
found that the behavioural activation of the green bugs, described
by the \emph{Source-Direction Movement Ratio}, has a nonmonotonic
behaviour as a function of the noise intensity $D$, with a maximum
at $D\approx 1.30 \cdot 10^{-5}$ $V^{2}$. This value represents the
optimal noise intensity since it maximizes the efficiency of the
sexual communication between individuals of \emph{N. viridula}.

This appears as the signature of the soft threshold stochastic
resonance (TSR)~\cite{Gree04}. By using a soft threshold model we
are able to compare through Figs.~\ref{fig:2}(b) and~\ref{fig:4} the
optimal noise intensities obtained in experiment and in numerical
simulations. The ensemble averaged cross-correlation coefficient
$<\rho>$ of the model~(\ref{soft}) shows a maximum at a noise
intensity value $D$ very close to the value that maximizes the SDM
ratio, observed in our experiments.

\vspace{0.5cm} \vspace{0.5cm} This work was supported by MUR and
INFM-CNISM. A.F. acknowledges the Marie Curie TOK grant under the
COCOS project (6th EU Framework Programme, contract No:
MTKD-CT-2004-517186).

\end{document}